\documentclass{moriond}
\usepackage{amsmath,amssymb}

\def\be{\begin{equation}}
\def\ee{\end{equation}}
\def\bea{\begin{eqnarray}}
\def\eea{\end{eqnarray}}

\def\sfrac#1#2{{\textstyle\frac{#1}{#2}}}

\newcommand{\Photo}{\includegraphics[height=35mm]{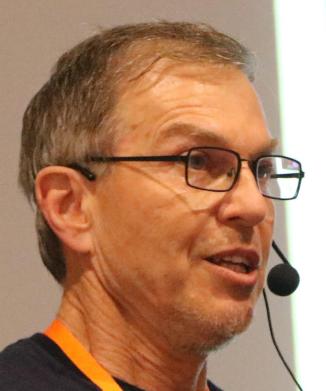}}

\begin{document}
\vspace*{4cm}
\title{Status of Dark Photons}

\author{James M.\ Cline}

\address{CERN, Theoretical Physics Department, Geneva, Switzerland,
and 
Department of Physics,\\ McGill University, 3600 rue
University, Montr\'eal, Qu\'ebec, Canada H3A2T8
}

\maketitle\abstracts{I review recent developments in the field of
dark photons---here taken to be U(1) gauge bosons with mass less than 
the $Z$---including both kinetically mixed vectors and those that couple to
anomaly-free U(1)'s.  Distinctions between Higgs and Stueckelberg
masses are highlighted, with discussion of swampland constraints,
UV completions, and new experimental search strategies.  
}

{\bf 1. Introduction.} As we have often been reminded, the possible
range of viable dark matter (DM)  masses spans many orders of magnitude, and
this is also true for DM in the form of U(1) vector gauge bosons.  The
heavier version is usually called a $Z'$ in the literature, whereas
lighter ones are called dark photons.  I take the the $Z^0$ mass to be
the upper boundary in this  review of dark photons, while below the
lower limit  $\sim 10^{-22}$\,eV, such particles would have too long a
de Broglie wavelength to function as dark matter in galaxies.\cite{Hui:2016ltb}
 Below this limit, dark photons could still play an
interesting role  in a larger dark sector, such as binding  dark
atoms  and giving rise to  millicharged dark matter (see Ref.\
\cite{Cline:2021itd} for a review).  Here I will not discuss massless
dark photons  except as an approximation to massive ones in the case
where the masses are negligible relative to relevant energy scales.

Hence, the dark photon could play the role of dark matter, dark force
mediator, or possibly both.  There are two ways in which it could
interact with the standard model: through kinetic mixing, which is
generically expected to be present, or by coupling to an anomaly-free
U(1) such $B-L$ or a difference between two lepton flavors.

We first recall how kinetic mixing works.\cite{Holdom:1985ag}
The Lagrangian for the SM U(1) gauge boson (either hypercharge, above
the electroweak symmetry breaking scale, or electromagnetism below it)
plus the dark one is
\be
{\cal L} = -\sfrac14 F_{\mu\nu} F^{\mu\nu} 
-\sfrac14 F'_{\mu\nu} F'^{\mu\nu} -\sfrac12\epsilon
F_{\mu\nu}F'^{\mu\nu} -\sfrac12 m_{A'}^2 A'_\mu A'^\mu\,.
\label{Lag}
\ee
If $m_{A'}$ is nonzero, the kinetic term is diagonalized (up to
$O(\epsilon^2)$ corrections) by $A_\mu\to A_\mu - \epsilon A'_\mu$.
Then the dark photon will couple to particles of charge $qe$ in the SM with
strength $\epsilon q e$.\footnote{\label{massesAp} If $m_{A'}=0$, one can instead
diagonalize the kinetic term using $A'_\mu\to A'_\mu + \epsilon
A_\mu$.  In this case, dark sector 
particles coupling to $A'$ become millicharged.  In the limit of
small $m_{A'}$, where $A'$ can be approximated as massless, for example
in a stellar environment, the dark Higgs looks
millicharged, which leads to strong constraints.}
It is interesting to notice that citations of Holdom's seminal paper
started to rise steeply in 2008, after the appearance of Refs.\
\cite{Arkani-Hamed:2008hhe,Pospelov:2007mp}, that highlighted the importance of kinetic mixing in extended
dark sectors.

\begin{figure}[t]
\centerline{\includegraphics[width=0.5\linewidth]{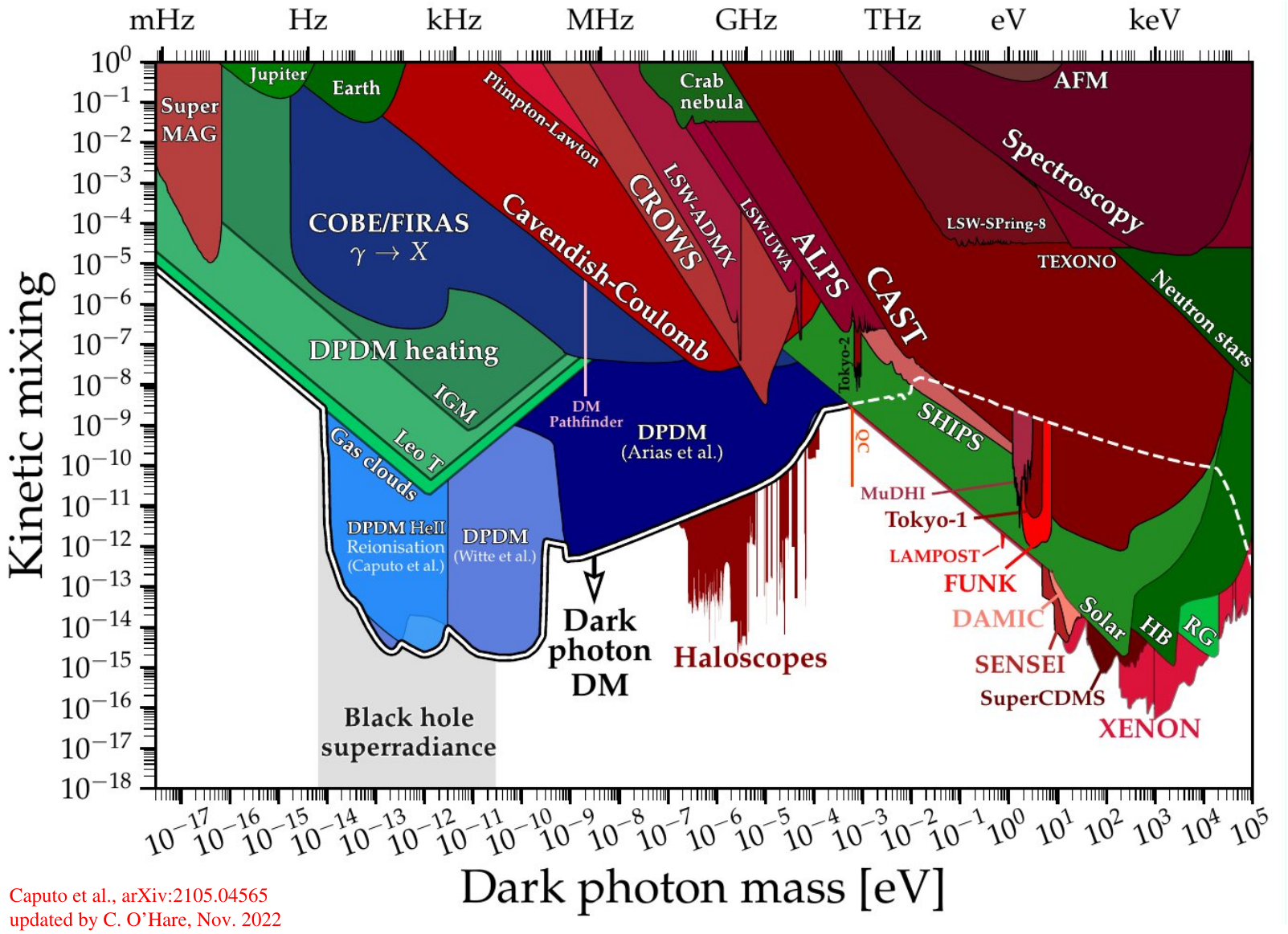}
\includegraphics[width=0.5\linewidth]{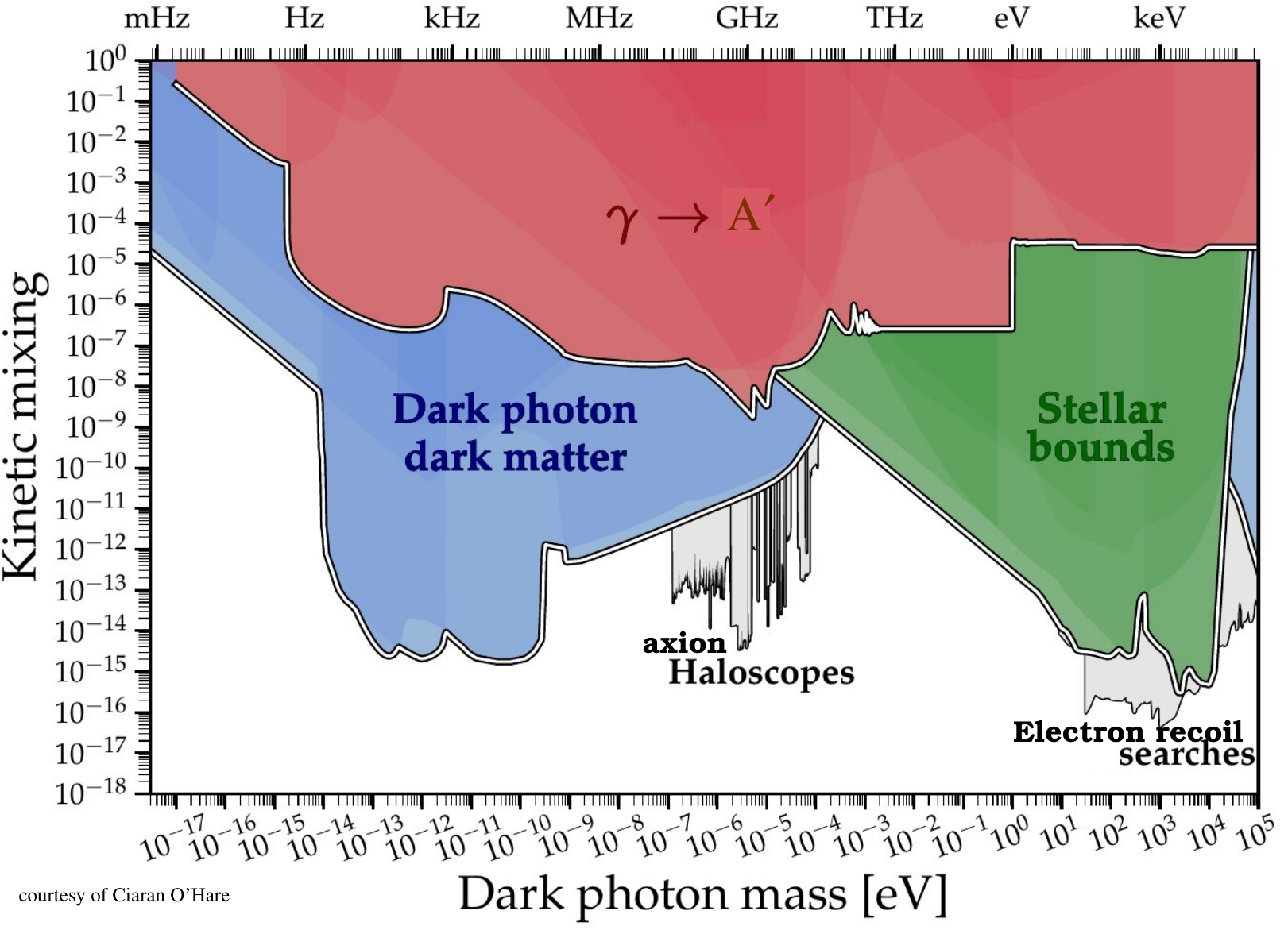}}
\caption[]{Left: recent limits on kinetic mixing versus mass, from
Ref.\ \cite{Caputo:2021eaa}, with updates from C.\ O'Hare. Right:
how the constraints are grouped by  categories.  Figure courtesy of
C.\ O'Hare.}
\label{fig:1}
\end{figure}

{\bf 2. Constraints on kinetic mixing.}
Experimental constraints on $\epsilon$ versus $m_{A'}$ have evolved
dramatically since 2008, when there were just a handful of limits
\cite{Ahlers:2008qc}.  Fig.\ \ref{fig:1} shows the current bounds
individually (left) and as classes (right).  The red region overlaps
with laboratory limits: $1/r^2$ force law tests,
light-shining-through-walls, helioscopes, axion haloscopes, dark
matter direct detectors, and a reactor neutrino experiment.  The
strongest current limit $\epsilon \lesssim 10^{-16}$ at
$m_{A'}\sim$ keV comes from $A'$ absorption by atoms in the 
XENON1T experiment.\cite{XENON:2019gfn}

The red region also includes 
constraints from SM photons resonantly converting to $A'$, 
due to the photon plasma mass matching $m_{A'}$ in the cosmic microwave
background, which would distort the cosmic microwave background, CMB (COBE/FIRAS constraint).  The
green region arises from $\gamma\to A'$ in stars, which would cause
the stars to cool faster than observed.  The foregoing constraints exist
independently of whether $A'$ is the DM.  If it is the DM, then the
blue regions are constrained by $A'\to\gamma$ causing reionization of 
helium in interstellar gas clouds. \cite{Caputo:2021eaa} 

Axion haloscopes are an example of experiments designed for different
physics whose limits can be reinterpreted in terms of
 dark photon detection.\cite{Arias:2012az,Ghosh:2021ard}  When the
cavity frequency matches $m_{A'}$, there are resonant photon-$A'$
oscillations, without the need for an external magnetic field.
 Recently it was shown
that radio telescopes can be sensitive to conversion of $A'\to\gamma$
in the dish leading to a monochromatic signal.  Assuming $A'$ is the
DM, LOFAR data rule out $\epsilon\gtrsim 10^{-13}$ for $m_{A'}\sim
10^{-7}\,$eV,\cite{An:2023wij} and FAST excludes $\epsilon \gtrsim 10^{-12}$ for
$m_{A'}\sim 10^{-5}\,$eV.\cite{An:2022hhb}  These limits are projected to
improve, especially with future SKA data.  A conceptually similar constraint
is shown to arise in the optical from the James Webb Space Telescope,\cite{An:2024kls}
 leading to $\epsilon\lesssim 10^{-11}$ for
$m_{A'}\sim (0.05-2)$\,eV.

\begin{figure}[t]
\centerline{\includegraphics[width=0.5\linewidth]{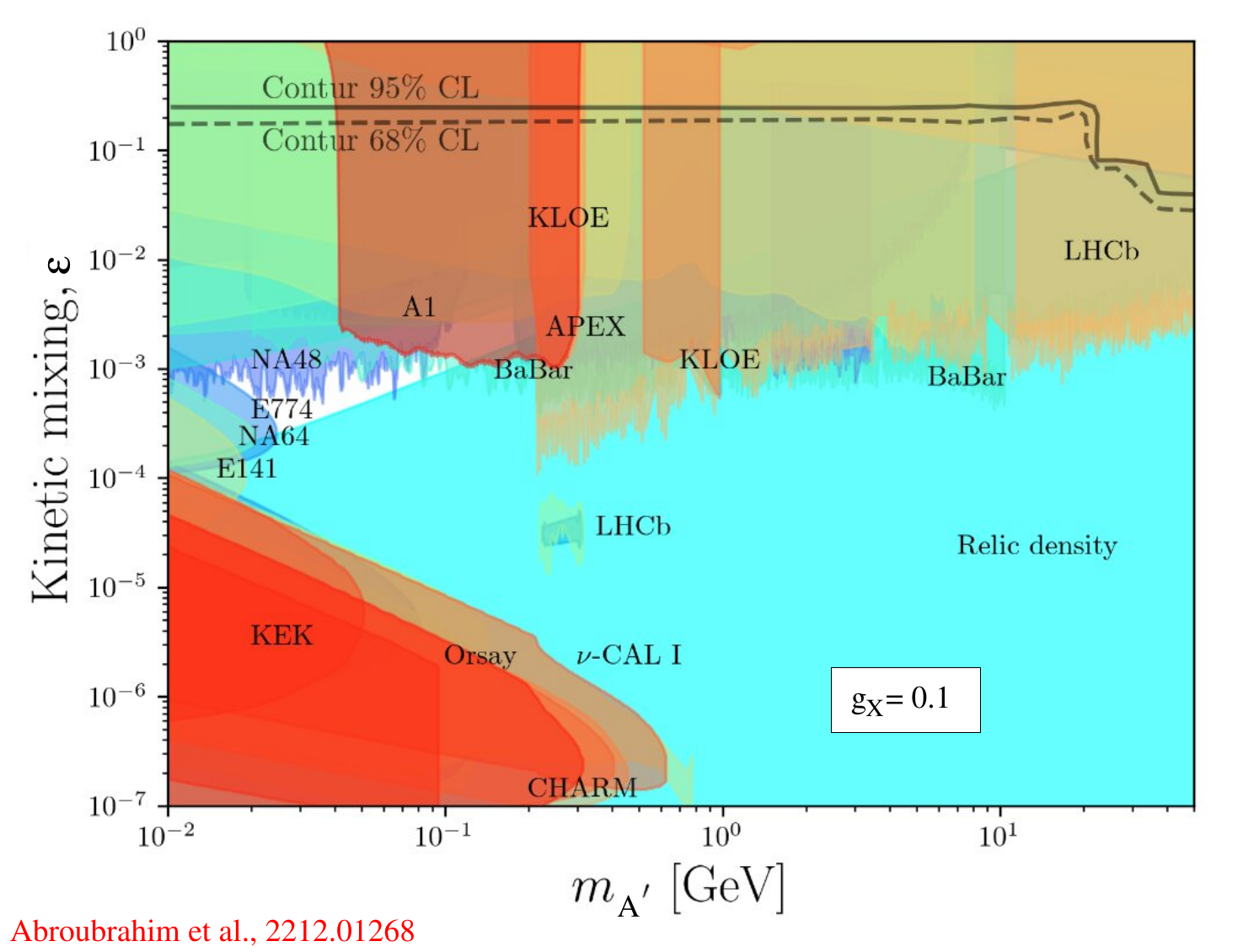}
\includegraphics[width=0.5\linewidth]{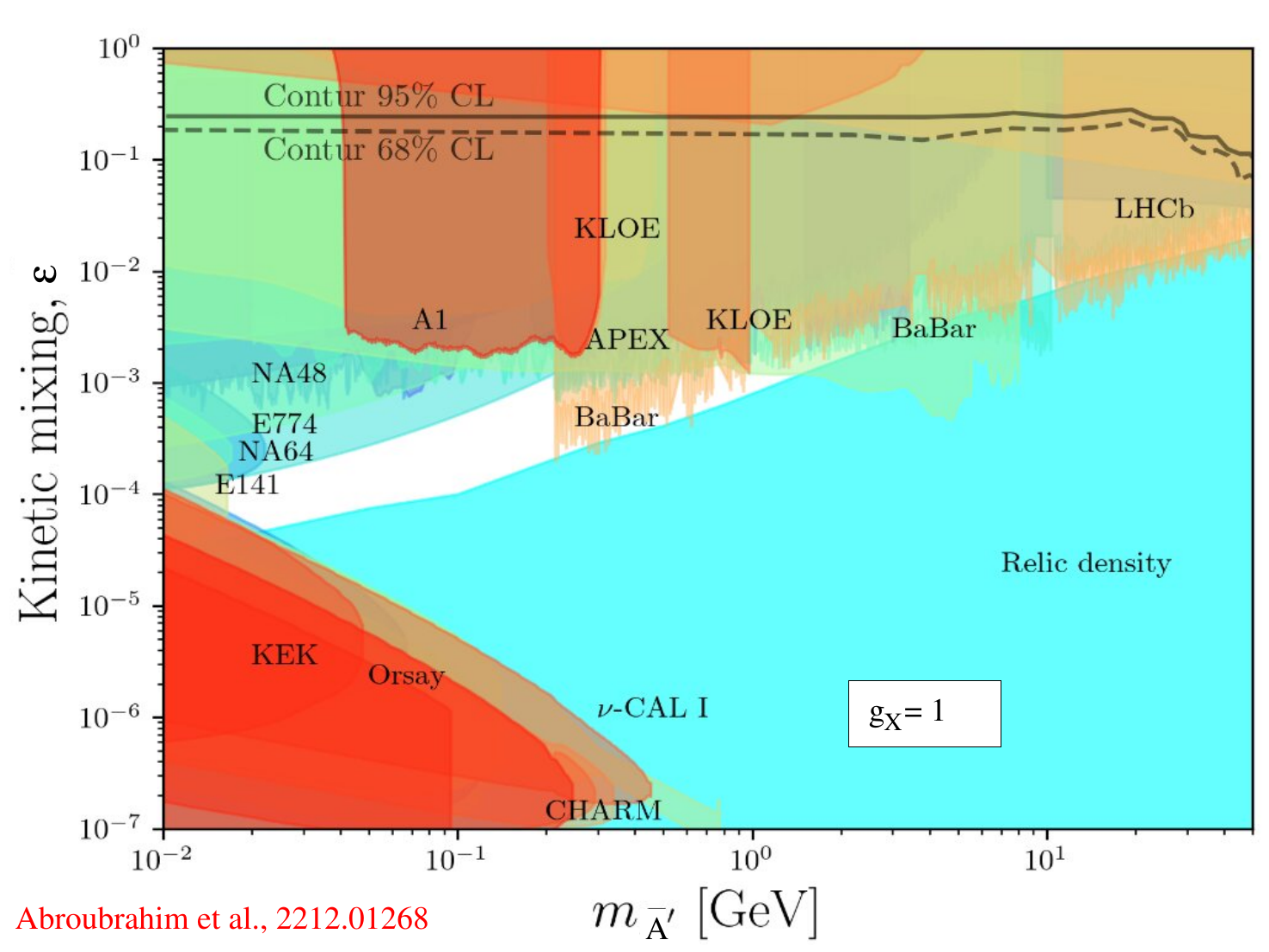}}
\centerline{\includegraphics[width=0.5\linewidth]{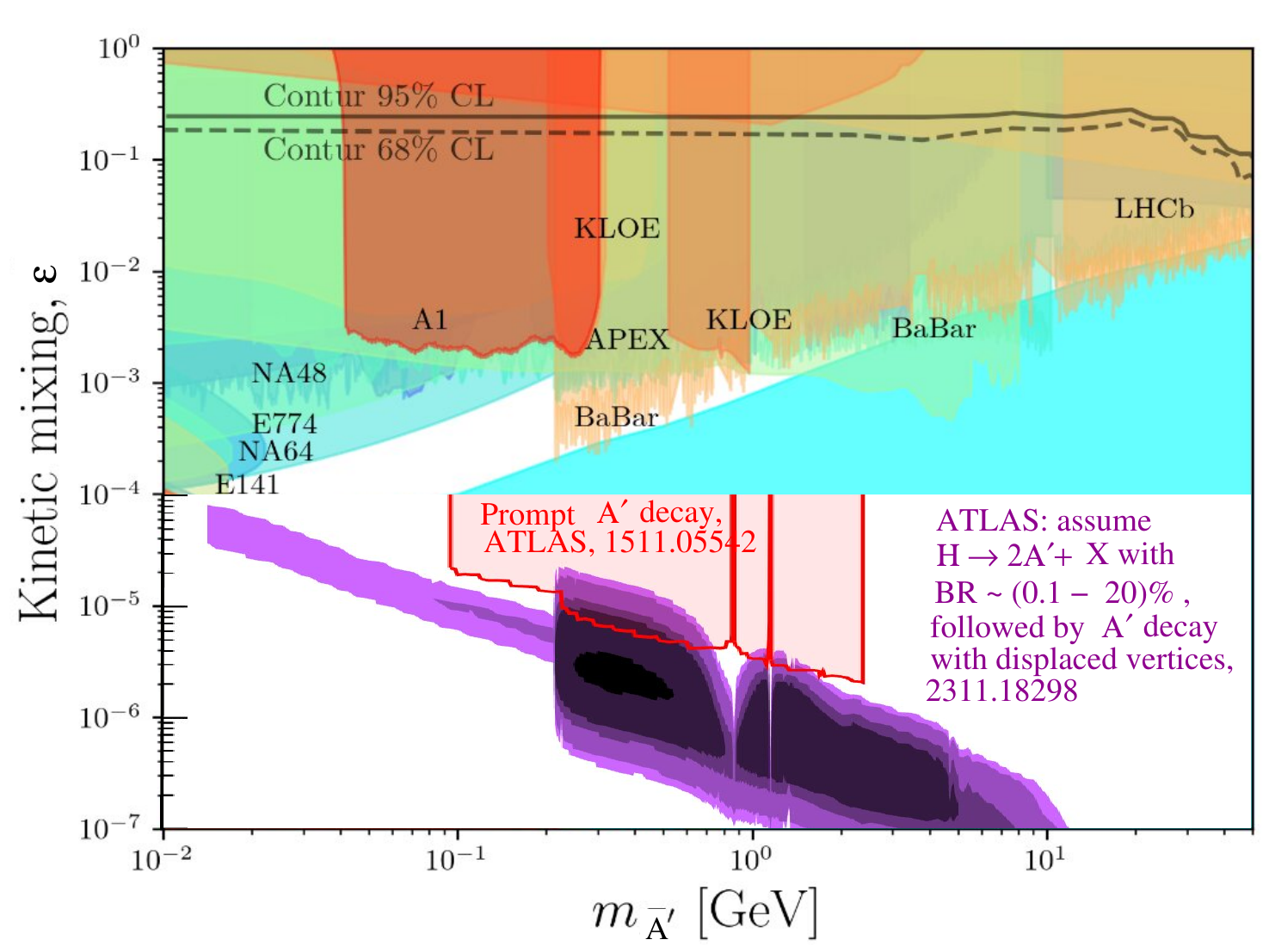}
\includegraphics[width=0.5\linewidth]{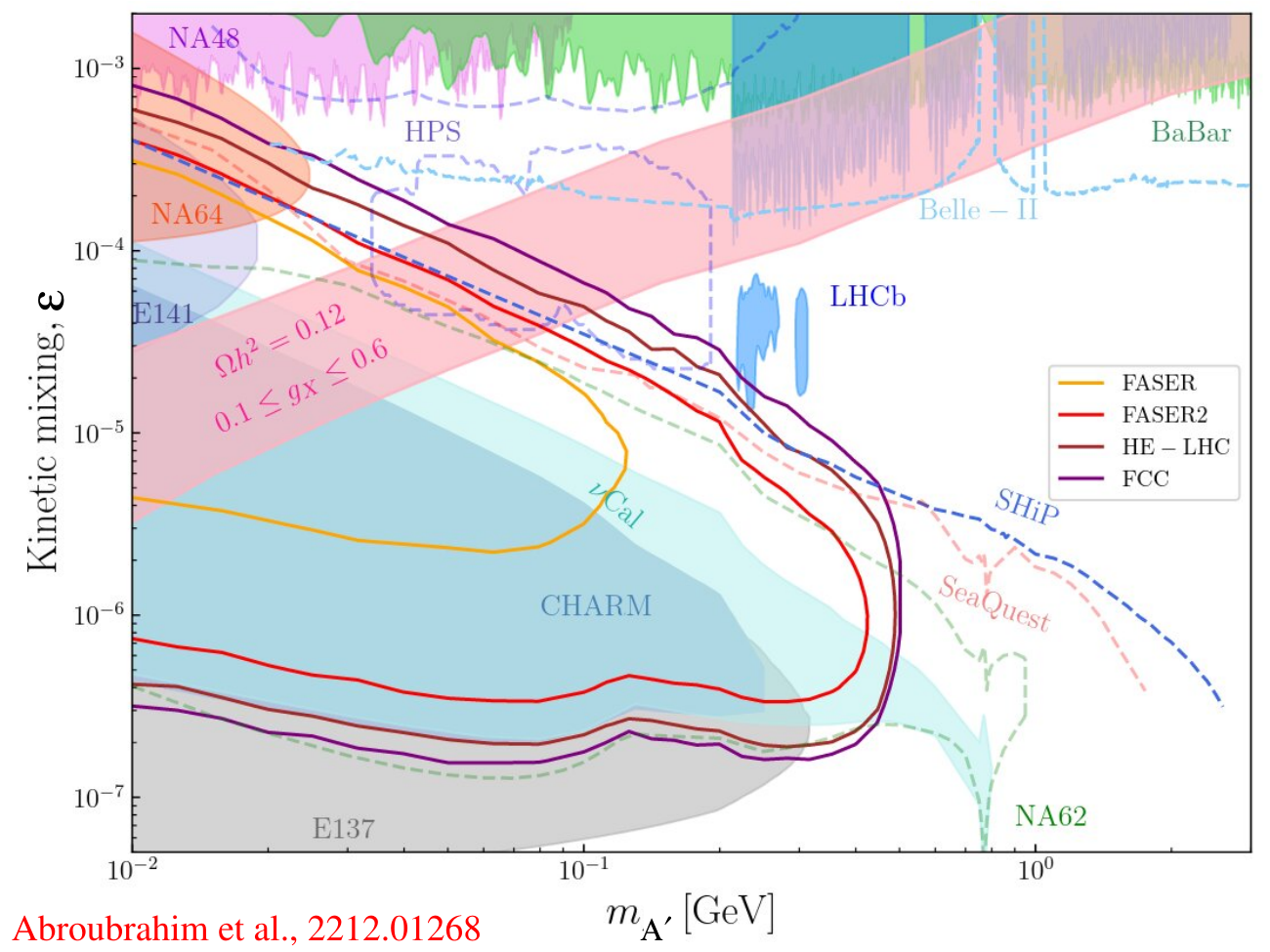}}
\caption[]{Top row: laboratory limits on kinetic mixing versus mass, 
with $A'$ decaying to lighter DM with coupling $g_X=0.1$ (left) or
1 (right), from
Ref.\ \cite{Aboubrahim:2022qln}.  Cyan region assumes DM freeze-in via
kinetic mixing. Bottom row: (left) ATLAS constraints from Higgs decay to dark
fermions, which then decay to $A'$ plus DM;\cite{ATLAS:2023cjw}
(right) anticipated sensitivity of future experiments for the
heavy-$A'$ parameter space.\cite{Aboubrahim:2022qln} }
\label{fig:2}
\end{figure}

So far we discussed light dark photons, $m_{A'} < \,$MeV.  Different
kinds of constraints apply for heavier $A'$, notably from an array of 
beam-dump and collider experiments.  Fig.\ \ref{fig:2} (top row)
illustrates these limits in the case where $A'$ couples to DM $\chi$
of mass $m_\chi = m_{A'}/4$ and coupling $g_\chi = 0.1$ or 1.  The
laboratory constraints depend on $g_\chi$ because it increases the
invisible width of $A'$.  However they generally resemble those
arising from purely visible decays of $A'$.  (The cyan regions are
disfavored by assuming freeze-in of DM through its interactions with
$A'$.) 

Model-dependent bounds have been obtained at LHC, illustrated in 
Fig.\ \ref{fig:2} (bottom left).  If the SM Higgs has a significant
branching ratio into dark fermions, which decay to $A'$ plus DM,
the subsequent decays of $A'\to \mu^+\mu^-$ yield strong limits,
depending on whether they are prompt or displaced.\cite{ATLAS:2023cjw}
  Reach of existing and proposed experiments to
further probe the parameter space is shown in the bottom right panel.
These include HPS, FASER, SHiP, SeaQuest, HE-LHC, and FCC.

\begin{figure}[t]
\centerline{\raisebox{3mm}{\includegraphics[width=0.5\linewidth]{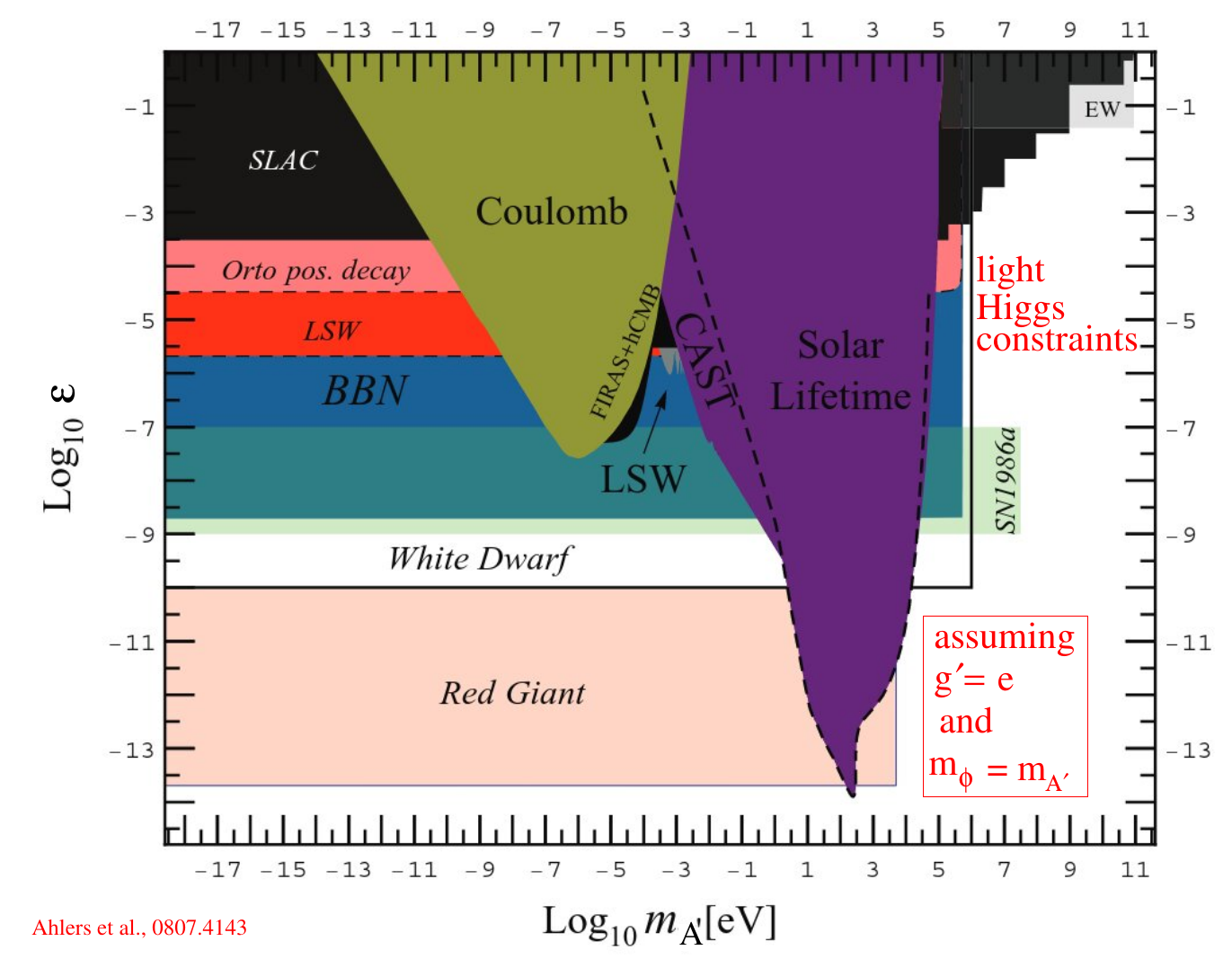}}
\includegraphics[width=0.5\linewidth]{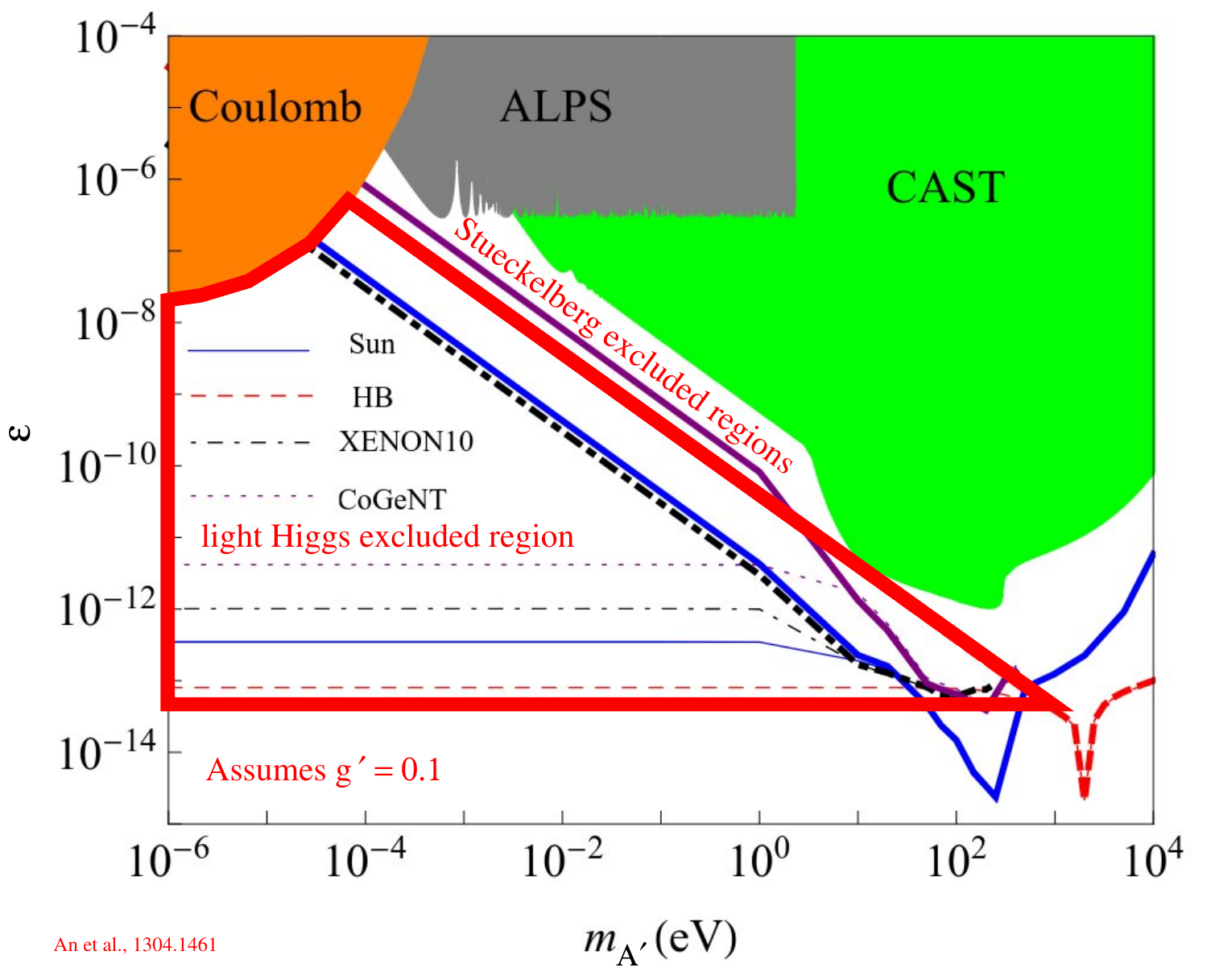}}
\caption[]{Left: enlarged exclusion regions for a light dark Higgs
from stellar cooling, SN1987a, and other effects.\cite{Ahlers:2008qc}
Right: similar constraints, also including $h'$ production in DM
detection experiments.\cite{An:2013yua} }
\label{fig:3}
\end{figure}

{\bf 3. Higgs or Stueckelberg mass?}
In presenting the previous bounds, it was implicitly assumed that the 
$A'$ gets its mass by the Stueckelberg ``mechanism,'' which
essentially means putting in a bare mass term as in Eq.\ (\ref{Lag})
with no other degrees of freedom.  One can restore gauge invariance by
introducing a fictitious field $\theta$ and rewriting the mass term
as 
\vskip-0.3cm
\be
\sfrac12 m_{A'}^2(A'^\mu -\partial^\mu\theta)^2\,,
\label{stueck}
\ee but this by
itself has no additional physical significance.  On the other hand,
if $A'$ gets its mass from a dark scalar $h'$ by the Higgs mechanism, one
might expect $m_{h'}\sim m_{A'}$.  If this is the case, then $h'$
behaves like a light millicharged particle, if $m_{A'}$ is small
compared to the energy scale of interest. There is a vertex $g' m_{A'}
h' A'^2$ which leads to $2\epsilon g' m_{A'}A'^\mu A_\mu$ in
conjunction with kinetic mixing.  This yields much stronger
constraints on $\epsilon$, ruling out large regions that were allowed
for the Stueckelberg case, as shown in Fig.\ \ref{fig:3}.  For example,
stellar cooling is accelerated by Brehmsstrahlung of $h'$ from
photons.\cite{Ahlers:2008qc} Further limits arise from production
of $h'$ in DM direct detection experiments  through
ionization of atoms in the detector by $A'$ produced in the sun.\cite{An:2013yua}  
These limits however are
model-dependent, because of the assumption $m_{h'}\sim m_{A'}$ and
$g'\sim e$.  

Typically, phenomenological studies assume that there is no analog of
the Higgs field corresponding to the Stueckelberg mechanism, but in
consistent UV completions, namely string theory, the field $\theta$ in
(\ref{stueck}) can be thought of  as an axion-like degree of freedom,
that has an accompanying radial excitation $\phi$.  Unlike the Higgs
mechanism, the $A'$ mass arises without $\phi$ getting a VEV, and the
limit $\phi\to 0$ in field space is singular because of a kinetic term
of the form $(\partial\ln\phi)^2$.\cite{Reece:2018zvv} In contrast
to Higgs masses, the limit $m_{A'}\to 0$ is thus inconsistent in the
Stueckelberg case.  

Light Stueckelberg  masses can naturally arise in large volume
string compactifications, but there is a lower limit on the mass,
which is related to the dimensionless volume ${\cal V}$ of the compactification
and thereby the string scale $m_s$ and the Planck mass, through $m_P^2 = (4\pi/g_s^2){\cal
V} m_s^2$.  The $A'$ mass is of order\cite{Goodsell:2009xc}
$m_{A'}\sim g_s^{3/2}(m_P/{\cal V})\gtrsim g_s^{3/2} {\rm\,
eV}\gtrsim 1{\,\rm eV}$,
using $m_s\gtrsim 1\,$TeV (from collider limits), hence ${\cal
V}\lesssim 10^{27}$.  

{\bf 4. Origin of kinetic mixing.}  The kinetic mixing operator
$F_{\mu\nu}F'^{\mu\nu}$ has dimension 4, so its coefficient $\epsilon$
can be put in ``by hand.''   But usually one imagines that it gets
generated at some high scale by integrating out a heavy particle that
carries both U(1)$'$ and SM U(1) charges.  In that case, its natural
value is of order $e g'/(16\pi^2)$.  To reconcile this with the
XENON1T limit would require $g'<10^{-11}$, precluding gauge
unification.  Otherwise, one could forbid matter charged under both
gauge groups, but this violates a strong version of the weak gravity
conjecture\cite{Arkani-Hamed:2006emk} (WGC, see below), which is supported  by known string
compactifications.

\begin{figure}[t]
\centerline{\includegraphics[width=\linewidth]{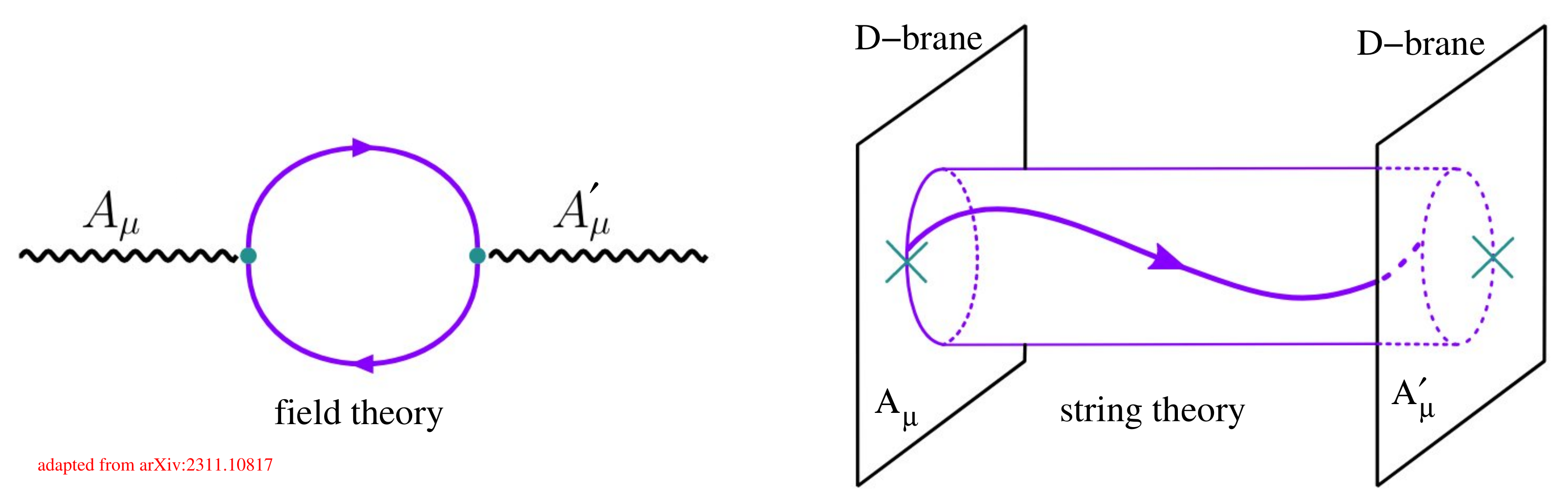}}
\caption[]{String theoretic origin of kinetic mixing between two
U(1)'s localized on different $D$ branes (right) versus field theory
diagram (left), from Ref.\ \cite{Hebecker:2023qwl}.}
\label{fig:4}
\end{figure}

If one ignores WGC and assumes $\epsilon=0$ the Planck scale, then
pure gravity plus Higgs mediation can generate $\epsilon$ through a
6-loop diagram, involving three graviton exchanges and one Higgs from
each (SM and dark) sector.\cite{Gherghetta:2019coi}  It is also
possible to evade the WCG constraints by having exchanges of several
heavy states in the loops whose contributions cancel each other.
This can happen naturally in string theory compactifications,\cite{Obied:2021zjc,Hebecker:2023qwl}
 where the interaction that
generates the kinetic mixing is a closed string exchange between two
$D$-branes, which host the respective $U(1)$ matter fields (Fig.\
\ref{fig:4}).  Small kinetic mixing can arise from large-volume
compactifications with ${\cal V} = (2\pi R)^6$ (in string length units
$\ell_s = 2\pi\sqrt{\alpha'}$), which are associated with a light
volume modulus of mass\cite{Conlon:2005ki} $m_{\cal V}\sim m_P{\cal V}^{-3/2}$.
  The kinetic mixing is related to $m{\cal V}$
by\cite{Hebecker:2023qwl,Goodsell:2009xc}
$\epsilon \gtrsim 10^{-16}\left(m_{\cal V}/ 2
m_h\right)^{8/9}$,
where $m_h$ is the SM Higgs mass.  To avoid overclosure of the
Universe by such moduli, they must be heavy enough to decay into two
Higgs bosons, hence this gives a lower bound on $\epsilon$ that is
close to the XENON1T constraint.

{\bf 5. DM Relic density from dark photons.} In constrast to axions
or other light spin-0 DM candidates, the misalignment mechanism
(where the field is stuck at a nontrivial value while the Hubble rate
$H$ exceeds its mass, and starts oscillating in its potential
subsequently) does
not easily account for a significant dark photon relic density.
This is because vector fields in a nontrivial gravitational background
couple to the Ricci scalar in such a way as to be diluted by the
expansion of the Universe.  One needs to introduce a nonminimal
coupling $R A'_\mu A'^\mu$  to counteract this effect,\cite{Arias:2012az,Golovnev:2008cf}
 but doing so leads to the
longitudinal polarization becoming ghost-like for some range of 
momenta.\cite{Karciauskas:2010as}  This can be overcome by modifying
the kinetic term by coupling it to a scalar field (the inflaton) 
via $f(\phi)F'_{\mu\nu}F'^{\mu\nu}$,\cite{Nakayama:2019rhg} but this involves significantly
 more model building than for a spin-0
particle, which only needs a quadratic potential for misalignment to
work.

Alternatively, a generic mechanism is through inflationary fluctuations of
$A'$, which like for any approximately massless field are of order
$H_I$, the Hubble rate during inflation.  Ref.\ \cite{Graham:2015rva} showed that
the desired relic density is achieved if 
$m_{A'} \cong 6\times 10^{-6}{\rm\,eV}\left(10^{14}\,\rm
	GeV/H_I\right)^4$.
Another relatively simple mechanism is to couple dark photons to an
axion via $-(\beta/f_a)a F'^{\mu\nu}\widetilde F'^{\mu\nu}$.\cite{Agrawal:2018vin,Co:2018lka}
  The axion can be produced by
misalignment; when it starts to oscillate, a tachyonic instability
quickly transfers energy from $a$ to $A'_\mu$ over a wide range of
axion masses and $A'$ masses, spanning $m_{A'}\sim(10^{-7}-10^7)$\, eV.
A similar mechanism works for dark Higgs oscillations producing 
$A'$ pairs by parametric resonance.\cite{Dror:2018pdh}

Dark photons can also serve as the mediator that gives rise to thermal
relic dark matter through the classic processes $\chi\bar\chi\to A'A'$
or $\chi\bar\chi\to A'^*\to f\bar f$.  However there are many other
possible production pathways that were explored in Ref.\ \cite{Hambye:2019dwd}.  For
example, kinetic mixing can be enhanced through resonant $A\to A'$
conversion at finite temperature, when $m_{A'}$ coincides with the
photon plasmon mass.

{\bf 6. Swampland constraints.}  The weak gravity conjecture has
strong evidence, and its variants are well motivated by string
theory/quantum gravity arguments.  The magnetic WCG imples a UV cutoff
$\Lambda > \sqrt{m_P m_{A'}/g'}$ for Stueckelberg masses.\cite{Reece:2018zvv}
Since $\Lambda > H_I$ during inflation for a
consistent description, the inflationary production mechanism for $A'$
becomes limited to masses $m_{A'}\gtrsim 0.3\,$eV, a very significant
restriction.   The WCG also demands that $\Lambda < g' m_P$, which
combined with the generic loop estimate for $\epsilon$ implies that
$\Lambda \lesssim (16\pi^2/e) \epsilon m_P\sim 100\,$TeV (using the
XENON constraint on $\epsilon$),\cite{Benakli:2020vng} suggesting new states would be accessible at the
Future Circular Collider. Similarly, combining the magnetic WGC with
the LHC bound $\Lambda \gtrsim 10\,$TeV and the loop estimate for
$\epsilon$ disfavors a large region of parameter space,\cite{Montero:2022jrc}
 including interesting values $m_{A'}\sim
10^{-11}$, $\epsilon\sim 10^{-7}$ where axion decays $a\to A'A'$
followed by $A'\to A$ resonant oscillations can boost the spectrum of
low-frequency CMB photons, suppressing the 21 cm signal as seen by the 
EDGES collaboration.\cite{Pospelov:2018kdh,Caputo:2020avy}

{\bf 7. Nonabelian kinetic mixing and Composite dark photons.}  It was
realized early on\cite{Arkani-Hamed:2008hhe} that the photon could kinetically mix with a
nonabelian gauge boson through the diagram\\
\centerline{\includegraphics[width=0.5\linewidth]{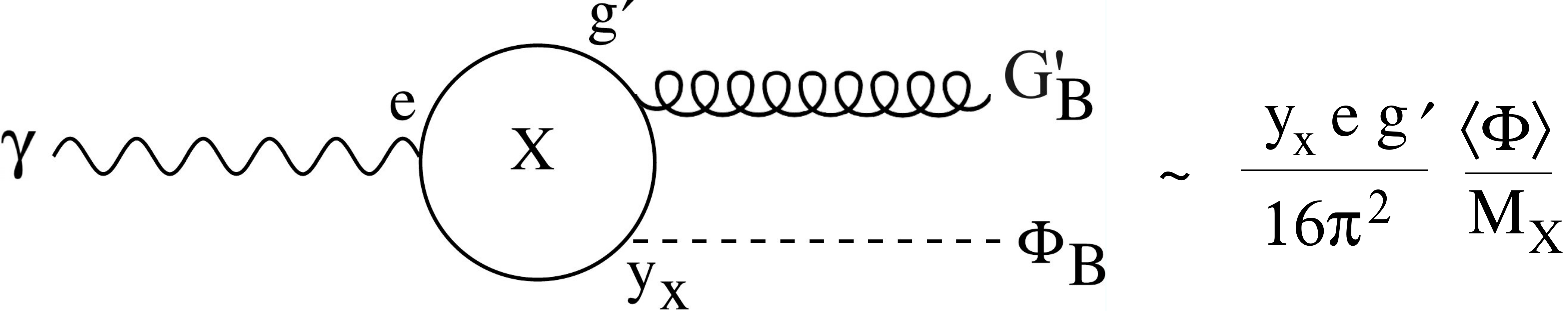}}
where $\Phi$ gets a VEV and breaks the gauge symmetry.  Then kinetic
mixing comes from a dimension-5 operator, giving 
another possibility for suppressing $\epsilon$, without needing small
$g'$.  It was pointed out\cite{Alonso-Alvarez:2023rjq} that this can also work in a confining
phase, where $\langle\Phi\rangle=0$; for example $\Phi$ binds with the
dark gluon to form a massive composite dark photon $\tilde A$, with
$\epsilon\sim \sqrt{\alpha\alpha'}\,y_X\Lambda /(4\pi m_X)$, where
$\Lambda$ is the dark confinement scale.  An interesting feature is
that the kinetic mixing $\epsilon$ can be correlated with $m_{\tilde
A} \sim \Lambda$ by their dependence on $\Lambda$ (note that
$\alpha'\sim 1/\ln\Lambda$ is not independent).  Dark baryons provide
a natural DM candidate in such models, and can have direct detection
signals by mediation of $\tilde A$.

{\bf 8. Gauging SM global symmetries.}  In addition to kinetic mixing,
$A'$ could couple directly to SM particles by gauging one of its
global symmetries.  The simplest possibilities involve anomaly-free
combinations, $B-L$, $L_e-L_\mu$, $L_e-L_\tau$ and $L_\mu-L_\tau$
since then no further matter content is required (except possibly
right-handed neutrinos).  If $\epsilon=0$ at
a high scale, then it is calculable at lower scales.\cite{Bauer:2018onh} For example 
$\epsilon \sim e g'\ln(m_\tau/m_\mu)/(4\pi^2)\sim 0.036\,g'$ for gauged $L_\mu-L_\tau$ 
at vanishing momentum transfer $q^2=0$; hence it makes a subdominant
contribution to observable processes.

\begin{figure}[t]
\centerline{\includegraphics[width=\linewidth]{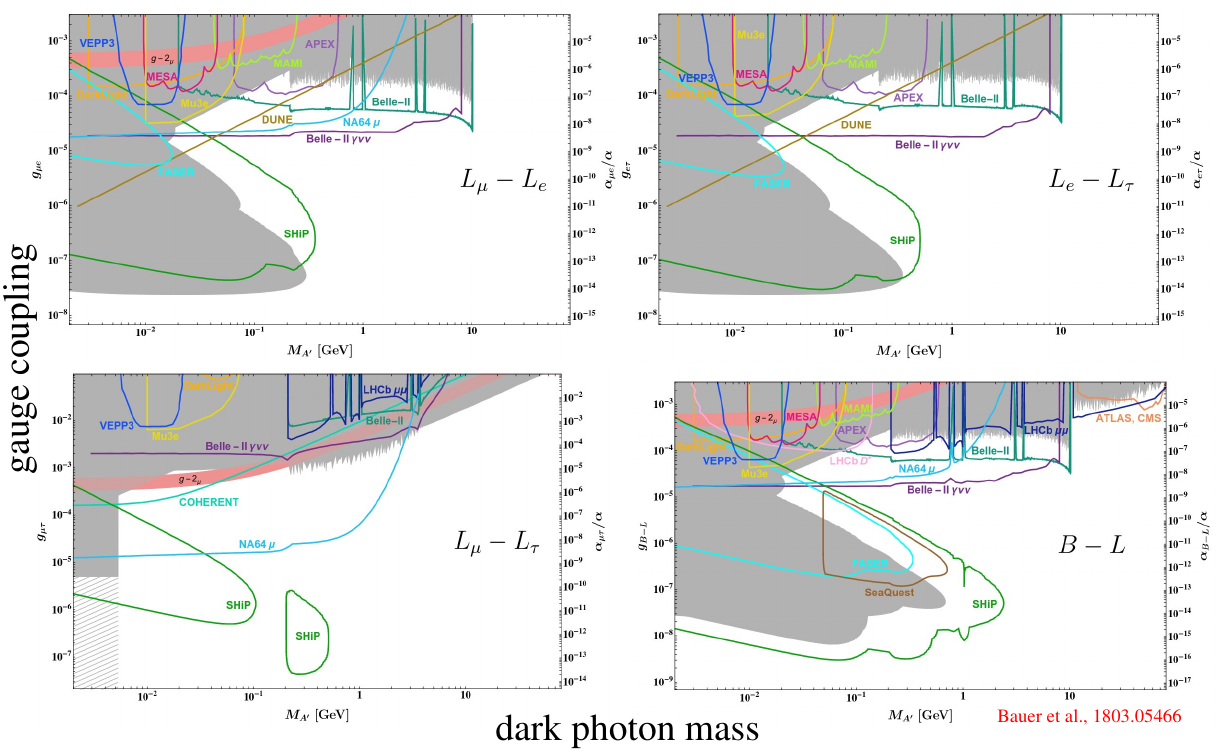}}
\caption[]{Constraints on gauge coupling versus $m_{A'}$ for the four
anomaly-free gauge groups, from Ref.\ \cite{Bauer:2018onh}.}
\label{fig:5}
\end{figure}

Constraints on $g'$ versus $m_{A'}$ for the gauged models are shown in
Fig.\ \ref{fig:5}.  They are most severe when $A'$ couples to
electrons, making $L_\mu-L_\tau$ a special case.  In fact, it is the
only one that can explain the muon $g-2$ anomaly (through one-loop
dressing of the muon vertex by $A'$ exchange), illustrated  by the
narrow pink band.  For the same reason, kinetically mixed models are
also ruled out for $(g-2)_\mu$ (see Ref.\ 
\cite{Mohlabeng:2019vrz} for a loophole).  

From a model-building
standpoint, it might seem arbitrary to gauge only $L_\mu-L_\tau$ as
opposed to other flavor combinations, but Ref.\ 
\cite{Alonso-Alvarez:2021ktn} showed that such a vector boson could be
just the lightest of eight such, coming from the gauging of SU(3)
lepton flavor, broken at the $(5-10)$\,TeV scale, except for one
scalar in the 6 representation with VEV $(20-200)$\,GeV to explain the
lightest vector mass.  The model predicts heavy neutral leptons at the
GeV scale, constrained by BBN, and that the lightest neutrino has mass
$m_{\nu_1}\gtrsim 10^{-4}$\,eV.  It can also address the
Cabbibo anomaly.

\begin{figure}[t]
\centerline{\includegraphics[width=0.53\linewidth]{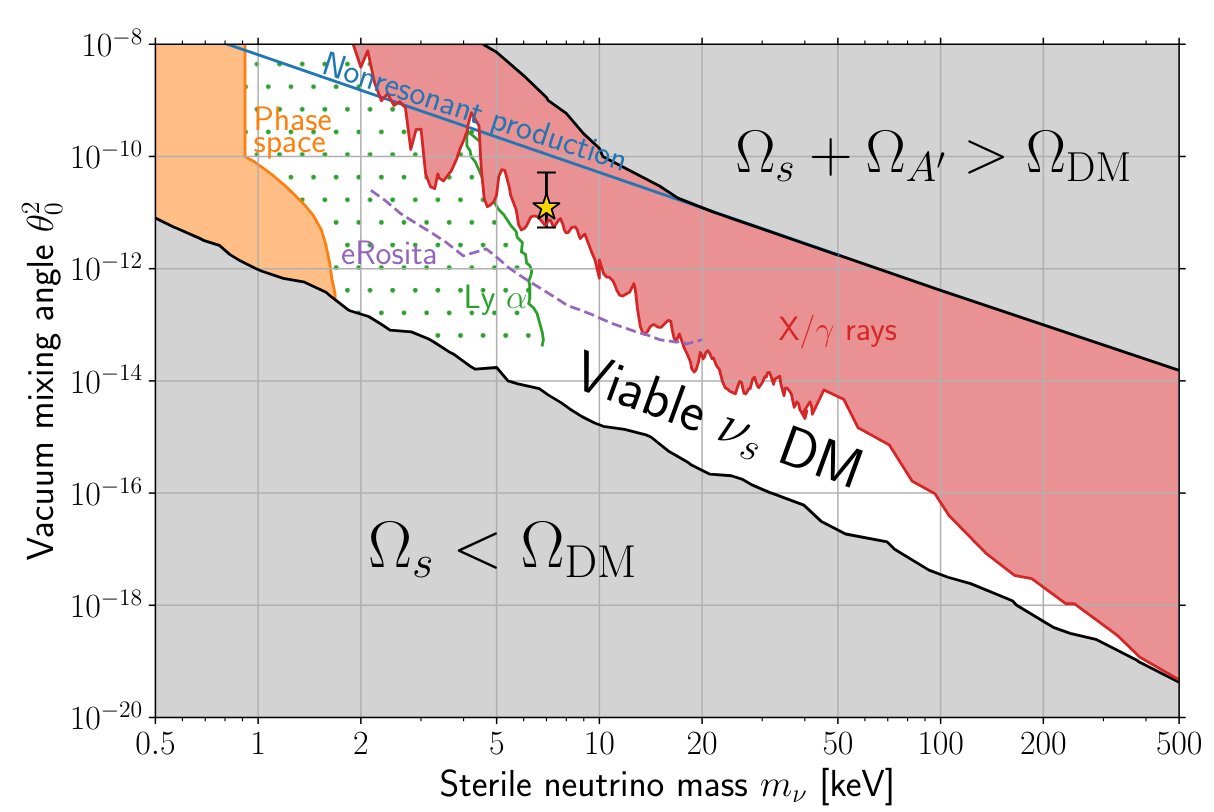}
\includegraphics[width=0.47\linewidth]{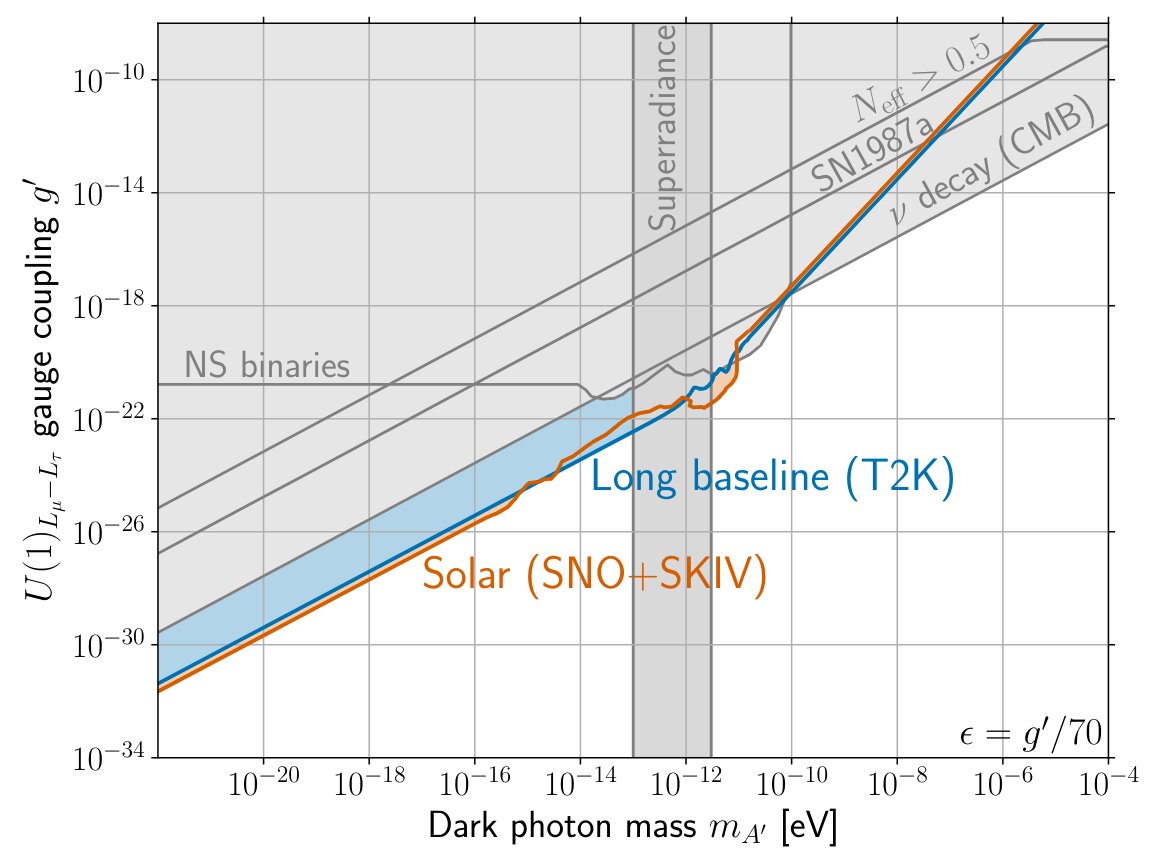}}
\caption[]{Left: re-opened region of active-sterile neutrino mixing
angle versus sterile $\nu$ mass, enabled by $L_\mu-L_\tau$ dark photon
dark matter.\cite{Alonso-Alvarez:2021pgy}   Right: constraints on the
$L_\mu-L_\tau$ gauge coupling versus $m_{A'}$ from distortions of
long baseline and solar neutrino oscillations by $L_\mu-L_\tau$ dark photon DM.\cite{Alonso-Alvarez:2023tii}}
\label{fig:6}
\end{figure}

The $L_\mu-L_\tau$ dark photon can have interesting implication for
neutrino physics if it is the dark matter. One can treat it as an
oscillating condensate which gives a time-dependent effective mass to $\nu_\mu$ and
$\nu_\tau$, with potential impact on neutrino oscillations.  This can
re-open parameter space for sterile neutrino dark matter, by making 
$\nu_s$-$\nu_\mu$ oscillations resonant and thereby making $\nu_s$
production efficient at smaller mixing angles than ordinarily
required;\cite{Alonso-Alvarez:2021pgy} this is illustrated in Fig.\ \ref{fig:6} (left). 
Similarly, this effect can impact oscillations between active neutrino
flavors,\cite{Brdar:2017kbt,Brzeminski:2022rkf,Alonso-Alvarez:2023tii} giving
rise to leading constraints on the coupling, as shown in Fig.\
\ref{fig:6} (right).  It was recently shown that $B-L$ dark photons can be
probed through dark matter-neutrino scattering if the DM carries
$B-L$, using IceCube observations of neutrinos from the active galaxy
NGC 1068.\cite{Cline:2023tkp}  Dark matter self-interactions mediated
by dark photons might also play a role in accelerating the mergers of supermassive
black holes,\cite{Alonso-Alvarez:2024gdz} to overcome the
long-standing ``final parsec'' problem.

{\bf Acknowledgments.}
This work was supported by the Natural Sciences and Engineering
Research Council (NSERC) of Canada.  I thank 
J.\ Jaeckel, C.\ O'Hare, M.\ Tytgat and S.\ Witte for helpful discussions, and  the CERN theory
department for its generous hospitality.

\end{document}